\documentclass[12pt]{article}
\usepackage[margin=0.7in]{geometry}
\usepackage[font=small,labelfont=bf]{caption}
\usepackage[numbers,square,sort&compress]{natbib}
\usepackage{authblk}
\usepackage{graphicx, amsmath}
\usepackage[latin1]{inputenc}
\usepackage[dvipsnames]{xcolor}
\usepackage{rotating}
\usepackage{xspace} % Sensible space treatment at end of simple macros

\usepackage[section]{placeins}
\makeatletter \renewcommand{\@biblabel}[1]{#1.} \makeatother
\graphicspath{%
{figs/}%
}
\begin{document}

\title{Quantum Thermodynamics of a Quantum Sized AdS Black Hole}

\author[1,2]{Behnam Pourhassan}
\author[1]{Mahdi Atashi}
\author[3]{Houcine Aounallah}
\author[2,4]{Salman Sajad Wani}
\author[2,5,6]{Mir Faizal}
\author[7]{Barun Majumder*}

\affil[1]{School of Physics, Damghan University,  Damghan 3671641167   Iran}
\affil[2]{Canadian Quantum Research Center, 204-3002 32 Avenue Vernon, British Columbia V1T 2L7 Canada}
\affil[3]{Department of Science and Technology, Larbi Tebessi University, Tebessa  12000   Algeria}
\affil[4]{Physics Engineering Department, ITU Ayazaga Campus 34469, Maslak, Istanbul}
\affil[5]{Department of Physics and Astronomy, University of Lethbridge, Lethbridge, Alberta, T1K 3M4 Canada }
\affil[6]{Irving K. Barber School of Arts and Sciences, University of British Columbia, Kelowna, British Columbia, V1V 1V7 Canada}
\affil[7]{University of Tennessee, Knoxville, Tennessee 37916, USA}
\date{}
\maketitle

\begin{abstract}
 In this paper, we investigate the effects of non-perturbative quantum gravitational corrections on a quantum sized AdS black hole. 
 It will be observed that these  non-perturbative quantum gravitational corrections modify the stability of this black hole. We will use the   non-equilibrium quantum thermodynamics to investigate the evaporation of this black hole between two states. We will analyze  the effects of   non-perturbative quantum gravitational corrections on this non-equilibrium quantum thermodynamics. We will explicitly obtain the  quantum work distribution for this black hole, as it evaporates between two states.  It will be observed that this  quantum work distribution is  modified due to non-perturbative quantum gravitational corrections. 
\vspace{0.5cm}
\end{abstract}

\section{Introduction}
%%%%%%%%%%%%%%%%%%%%%%%%

It can be demonstrated using quantum field theory in curved spacetime that  black holes  emit a thermal radiation called Hawking radiation  \cite{1a,1b,1c, 1a0}.  The  temperature of the Hawking radiation is inversely proportional  to its surface gravity.  The black holes also   have an  entropy,  which scales with the area of the horizon. This  scaling property of the  entropy has led to the development  of the holographic principle  \cite{20,21}. As this analysis is  based on quantum field theory in curved spacetime,  it is a semi-classical  approximation. Thus, it is expected that the black hole thermodynamics will be corrected due to quantum gravitational corrections.  In fact, it has been explicitly demonstrated that the relation  between the  entropy and the area of a black hole would be  modified by  quantum gravitational  corrections \cite{2a,2b,2c,2d}. So, these quantum corrections to the structure of spacetime can produce thermal fluctuations \cite{dumb} to the thermodynamics of black holes. However,  even these  correction terms  are functions of the area, rather than the volume, and so the holographic principle can still be used to analyze them. In fact, as AdS/CFT is a concrete realization of the holographic principle, it has been used to obtain  such quantum gravitational corrections to the entropy of a black hole \cite{3a,3b,3c,3d}. It  is possible to use various different approaches to quantum gravity  to obtain corrections to the black hole thermodynamics \cite{3e,3f,3g,3h,3i}. The effects of such   quantum gravitational corrections on  rotating black hole in AdS spacetime \cite{q0}
charged black hole solution in Rastall theory \cite{q1,q2}, AdS black hole with a monopole \cite{q4}, Skyrmion black holes \cite{qq44} and black hole in a hyperscaling violation background \cite{q5} have been investigated. It was observed that these corrections    produce important modifications to the equilibrium thermodynamics of these black holes. 

The connection between quantum corrections to the structure of spacetime, and thermodynamics of black holes can be clearly seen from the Jacobson formalism  \cite{teda}. In this formalism,  Einstein equation can be derived from thermodynamical considerations \cite{teda}. It has also been observed using the Jacobson formalism, the  quantum fluctuations to the structure of  spacetime can be an obtained from  thermal fluctuations  \cite{1d}. At large scales, the quantum fluctuations can be neglected, and the geometry can be described by a classical geometry. As the  temperature of a black hole scales inversely with its mass, for large black holes, such  thermal fluctuations can also be neglected, and the system can be described by using  equilibrium thermodynamics. However, for sufficiently small black hole, the perturbative quantum corrections produce   thermal fluctuations to the equilibrium thermodynamics of the system  \cite{1e,1f,1g,1h,1i,1j}.  The effects of both the leading order thermal corrections   \cite{log1,log2,log3,log4}, next-to-the leading order thermal corrections   \cite{higher1,higher2} on  black hole thermodynamics have been investigated. However, such corrections only only at a scale, where the perturbative corrections to the equilibrium entropy are still valid. Now,  for a quantum sized  black hole, whose size is  comparable to the Planck scale,   this perturbative treatment is expected to breakdown. At such a small scale, we have to include non-perturbative quantum gravitational corrections.  It has been argued that the non-perturbative quantum gravitational corrections would modify the original equilibrium entropy of a black hole by an exponential function  \cite{3}.
In fact, it is known that the such corrections can also be obtained using supergravity functional integral  \cite{3aa,  ds12,  ds14}.
The effect of such corrections on the thermodynamical stability  of  spherical symmetric black holes has been investigated \cite{3cd}. It has also been demonstrated that such corrections modify the behavior of a Born-Infeld black hole in a spherical cavity \cite{3bb}. These non-perturbative corrections to black branes thermodynamics have been used to construct a quantum corrected geometry for black branes \cite{3ab}.

The   AdS black holes has been constructed as solutions to  supergravity approximation of  string theory \cite{4, 4a, 4b, 4c, 4d}.  
The thermodynamics of AdS black holes is important as it can be investigated using the  thermodynamics of the  conformal field  theory   dual  to it \cite{cft1, cft2}.
Thus, the corrections  to the thermodynamics of the dual theory  can be used to obtain the corrected thermodynamics of the AdS black hole  \cite{3a,3b,3c,3d}.  
In this paper, we will analyze the effects of    non-perturbative exponential corrections  \cite{3aa,   ds12,  ds14} on the  thermodynamics of an      AdS black hole \cite{5, 51, 52, 54}.  
It is known that in an AdS black hole the cosmological constant can be treated as the thermodynamic pressure, and   its thermodynamics can be studied in an  extended phase space  \cite{pv1, pv2, pv4, pv5}. It has been observed that this     extended phase space thermodynamics of  an AdS black hole gets modified due to perturbative quantum gravitational corrections \cite{pv6, pv7}.
Here, we will also analyze the effect of non-perturbative quantum corrections to the  thermodynamics of an AdS black hole in an extended phase space.

As these corrections are non-perturbative, they would correspond to non-equilibrium thermodynamics. Thus, we will use the formalism of non-equilibrium quantum thermodynamics \cite{mp6,mp7} to investigate the behavior of a quantum sized  AdS black  hole.
We will calculate the    quantum work distributions between two states of such a black hole using this non-equilibrium quantum thermodynamics \cite{mp8,mp9}. This quantum work in quantum thermodynamics is an analog of classical work, and is obtained using the Crooks fluctuation theorem \cite{6a}.  The   quantum work distributions between two states of  a system can be directly related to the free energies between them using the Jarzynski inequality  \cite{6b}. The  free energy of a  black hole can be obtained from standard black hole thermodynamics, and this can be used to obtain  quantum work distribution for a black hole \cite{6c, 6c1}. It may be noted that  the quantum work distribution is only significant at very small scales, and at such scales we cannot neglect the quantum gravitational corrections to the system. To properly analyze the quantum work distribution for AdS black holes, we need to use the non-perturbative quantum gravitationally  corrected free energies.  Thus, we have to use the    non-perturbative  quantum gravitational corrections \cite{3, 3aa,  ds12,  ds14}  to  obtain corrected  free energies for an AdS black hole.  Then  the  quantum work distribution for an evaporating   quantum sized  AdS black hole \cite{6c, 6c1} can be obtained using these quantum corrected free energies. It may be noted that the corrections to the   quantum work distribution from non-perturbative  quantum gravitational corrections have been obtained for a system of M2-M5 branes \cite{6b1}, and a quantum sized Myers-Perry black hole \cite{6b2}.  Here, we will use this formalism \cite{6b1, 6b2} to obtain the quantum work distribution for a quantum  sized  AdS black hole.

\section{Quantum Gravitational Corrections}
\begin{figure}[h!]
\begin{center}$
\begin{array}{cccc}
\includegraphics[width=70 mm]{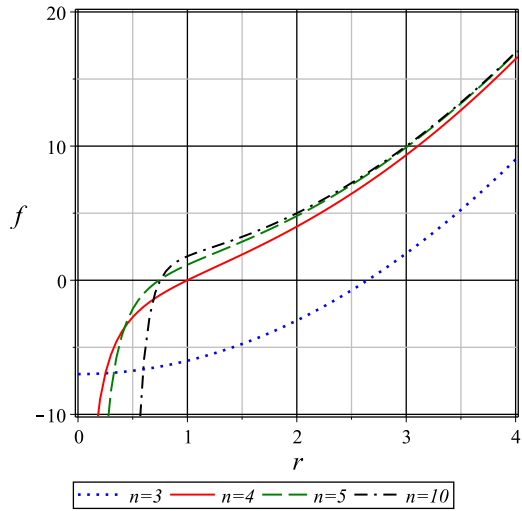}
\end{array}$
\end{center}
\caption{Horizon for a $n$-dimensional Schwarzschild-Tangherlini AdS black hole, for $M=1$ and $l=1$.}
\label{fig-f}
\end{figure}

In this section, we will analyze the quantum gravitational corrections  \cite{3, 3aa} to the thermodynamics of a $n$-dimensional Schwarzschild-Tangherlini AdS black hole \cite{5, 51, 52, 54}.
These quantum gravitational corrections will modify the original entropy of this AdS black hole \cite{3, 3aa,  ds12,  ds14}, and these modifications will in turn change the thermodynamic behavior of other thermodynamics quantities.
The metric of a $n$-dimensional Schwarzschild-Tangherlini AdS black hole can be written as \cite{5, 51, 52, 54}
\begin{equation}\label{metric}
ds^{2}=-f\left(r\right)dt^{2}+\frac{dr^{2}}{f\left(r\right)}+r^{2}d\Omega_{n-2}^{2},
\end{equation}
where $f(r)$ in the metric can be expressed as
\begin{equation}
f\left(r\right)=1-\frac{16\pi M}{\left(n-2\right)\omega r^{n-3}}+\frac{r^{2}}{l^{2}},
\end{equation}
with  $M$ as the ADM mass of the black hole,   $l$ as the AdS  radius, and  $\omega={2\pi^{\frac{n-1}{2}}}/{\Gamma\left(\frac{n-1}{2}\right)}$ as the area of a $n-1$-dimensional unit sphere.
In Fig. \ref{fig-f}, we investigate  the structure of the horizon for a $n$-dimensional Schwarzschild-Tangherlini AdS black hole. We observe   there is a positive real root for $f(r)=0$, which can be  denoted by $r_{h}$.

As the thermodynamics of black holes can be studied using  a conformal field theory, thus it is  possible to obtain  perturbative corrections to the thermodynamics of black holes from such  a conformal field theory   \cite{mi12, mi14,mi16, mi18}. The   modular invariance of the partition functions for a  conformal field theory \cite{mi12, mi14,mi16, mi18}, can be directly used to demonstrate that the corrections to the entropy of a black hole can be expressed   in terms of  the original equilibrium entropy of the black hole $S_0$ as $S_{\text{per}}   \sim   \ln (S_0)$
\cite{1e,1f,1g,1h,1i,1j}.
This is done by investigating the effects of    small fluctuations around the equilibrium on the original black hole entropy.  So,  using the   modular invariance of the  partition function of the conformal field theory \cite{mi12, mi14,mi16, mi18},   it is possible to express this corrected entropy as    $S(\beta) = a \beta^n + b \beta^m,$ where    $a, b , n, m, >0$ are constants.  At extremum for $S(\beta)$,   $\beta_0 = (nb/ma)^{1/m+n} = T^{-1}$ represents  the original  equilibrium temperature \cite{1e,1f,1g,1h}. The perturbative corrections can then be  expressed in terms of this equilibrium temperature, and the equilibrium entropy $S_0 = S(\beta)|_{\beta = \beta_0}$. These  perturbative corrections can explain the behavior of small black holes, which are still large enough to neglect the full non-perturbative quantum gravitational corrections.  However, for quantum sized  black holes, we have to consider  the effects of      non-perturbative  quantum gravitational corrections to the black hole entropy. These  non-perturbative  quantum gravitational corrections  to the black hole entropy again can be expressed in terms of the original equilibrium entropy of a black hole as  $S_{\text{non-per}}\sim \exp {-S_{0}}$ \cite{3, 3aa, 3bb, ds12,  ds14}.
As this form of the correction has been proposed to be universal   \cite{3, 3aa, 3bb, ds12,  ds14}, we will analyze the effects of such corrections on the thermodynamics of  an AdS black hole. We will specifically introduce a control parameter $\eta$ to control the  effects of such non-pertubative  corrections, and write the corrected entropy as   $ S=S_{0}+ \eta e^{-S_{0}}$.  This introduction of a control parameter  has been  motivated from the use of such a control parameter for  perturbative corrections \cite{1i,1j}.
Thus, using the original entropy of a $n$-dimensional Schwarzschild-Tangherlini AdS black hole, we can write the  corrected entropy of a quantum sized AdS black hole as
\begin{equation}\label{mod-ent}
S=\frac{\omega}{2}r_{h}^{n-2}+\eta \exp\left({-\frac{\omega}{2}r_{h}^{n-2}}\right)
\end{equation}
 It should be observed that this corrected entropy term  reduces  the original entropy of an  AdS black hole,  when the AdS black hole is large enough for the quantum gravitational corrections to be neglected. Only at very small scales, at which non-perturbative quantum gravitational corrections cannot be neglected,  we have to use this corrected AdS entropy.

\begin{figure}[h!]
\begin{center}$
\begin{array}{cccc}
\includegraphics[width=70 mm]{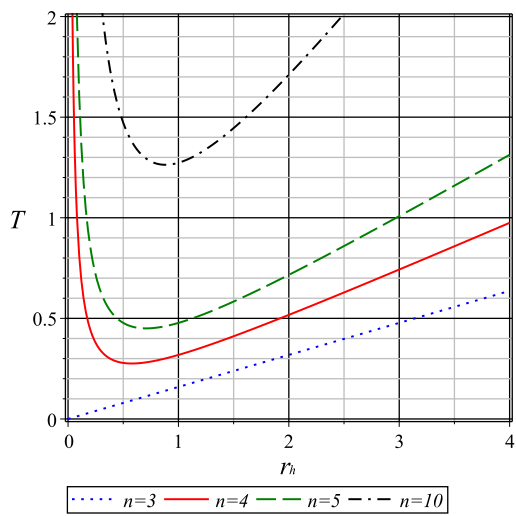}
\end{array}$
\end{center}
\caption{Temperature of $n$-dimensional Schwarzschild-Tangherlini AdS black hole for $l=1$.}
\label{fig-T}
\end{figure}

It is important to observe that these corrections are expressed in terms of the original temperature and original equilibrium entropy. Thus, we will use the original  temperature for this AdS black hole, which can be written as
\begin{equation}\label{T}
T=\frac{1}{4\pi}\left(\frac{d{f\left(r\right)}}{dr}\right)_{r=r_{h}}=\frac{(n-1)r_{h}^{2}+(n-3)l^{2}}{4\pi l^{2}r_{h}},
\end{equation}
where in the last equality we remove ADM mass using horizon radius.
Now for   $n=3$ and $n=5$, the horizon radius $r_{h}$, can be written as
\begin{eqnarray}
r_{h}&=&l\sqrt{\left(\frac{16\pi M-\omega}{\omega}\right)}; \hspace{46mm} n=3,\\
r_{h}&=&\sqrt{-\frac{l^{2}}{2}+\frac{\sqrt{\left(3l^{2}\omega\right)^{2}+12\left(16\pi Ml^{2}\right)}}{6\omega}};\hspace{20mm} n=5.
\end{eqnarray}
For these cases, the temperatures can be expressed as
\begin{eqnarray}
T&=&\frac{1}{2\pi l}\sqrt{\left(\frac{16\pi M-\omega}{\omega}\right)};\hspace{89mm} n=3, \\
T&=&\frac{1}{2\pi\sqrt{-\frac{l^{2}}{2}+\frac{\sqrt{\left(3l^{2}\omega\right)^{2}+12\left(16\pi Ml^{2}\right)}}{6\omega}}}
+\frac{1}{\pi l^{2}}\sqrt{-\frac{l^{2}}{2}+\frac{\sqrt{\left(3l^{2}\omega\right)^{2}+12\left(16\pi Ml^{2}\right)}}{6\omega}};\hspace{8mm} n=5.
\end{eqnarray}
In Fig. \ref{fig-T}, we plot the temperature in terms of the radius of the horizon. For $n\geq4$, we can observe  that there is a minimum value for the temperature. However,  for $n=3$ temperature is a linear function of  $r_{h}$. We can use this original temperature, along with the corrected entropy to analyze the effect of quantum gravitational corrections to on the thermodynamics of this   quantum sized AdS black hole.

\section{Stability}
\begin{figure}
\begin{center}$
\begin{array}{cccc}
\includegraphics[width=50 mm]{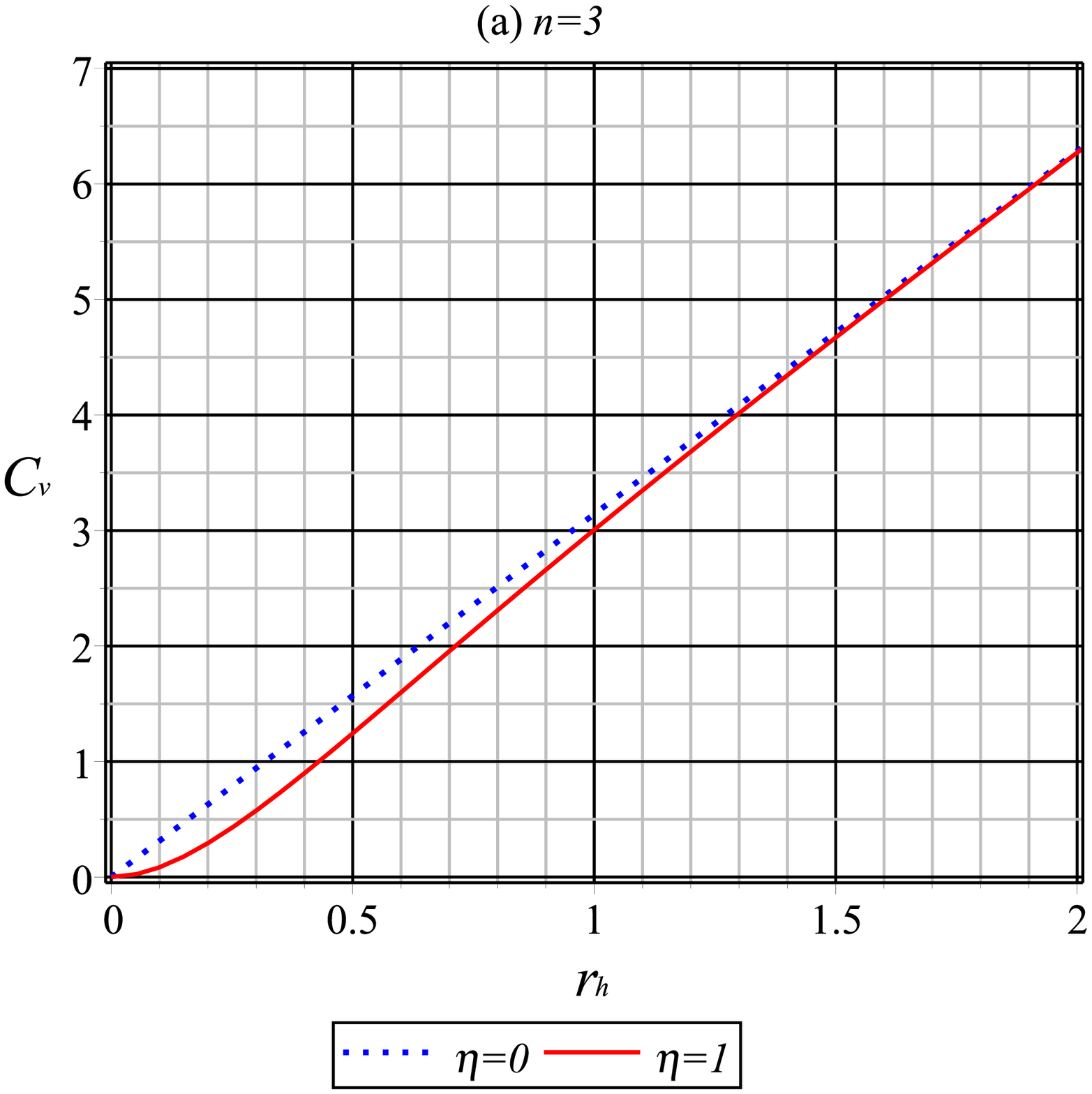}\includegraphics[width=50 mm]{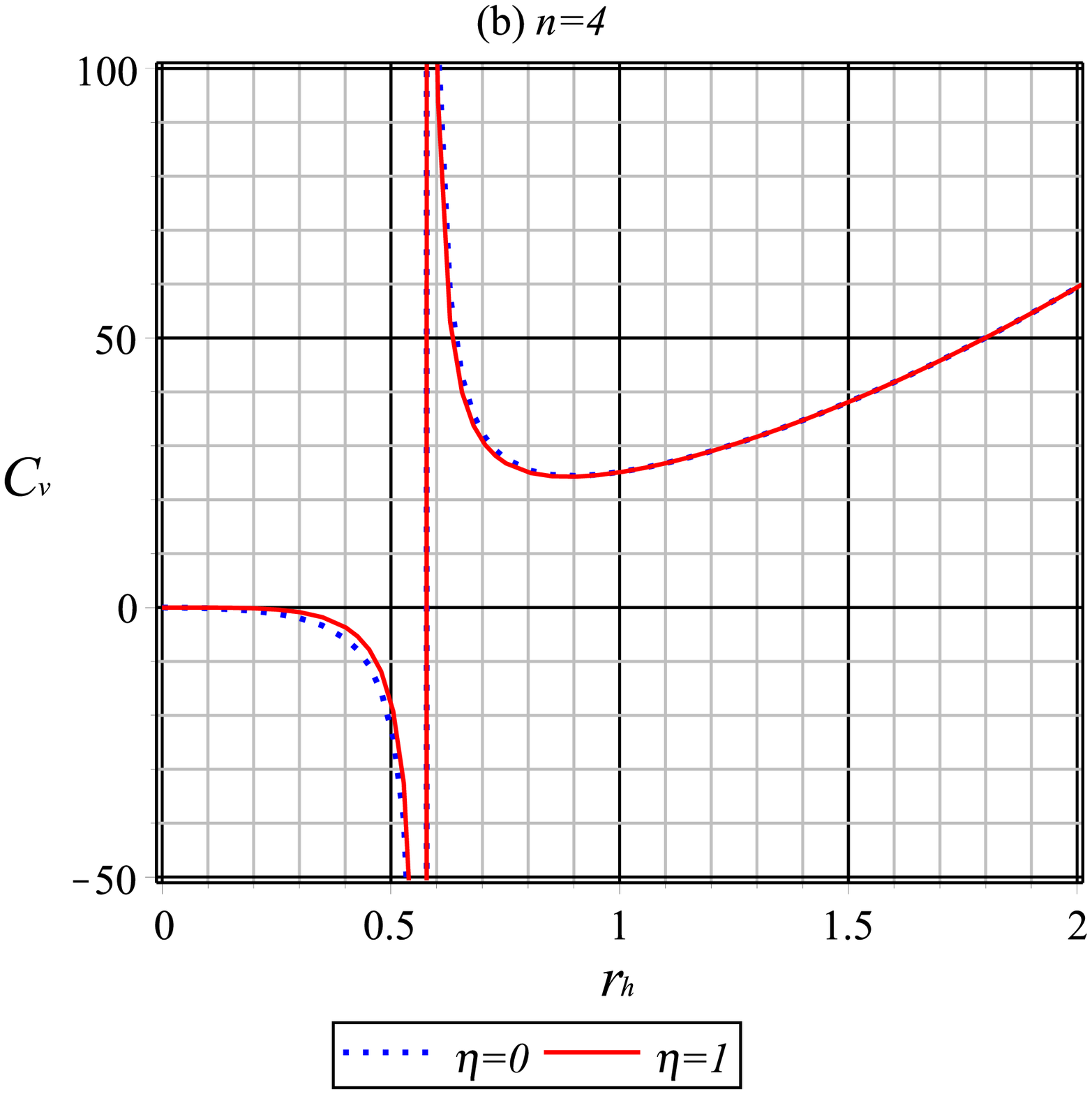}\\
\includegraphics[width=50 mm]{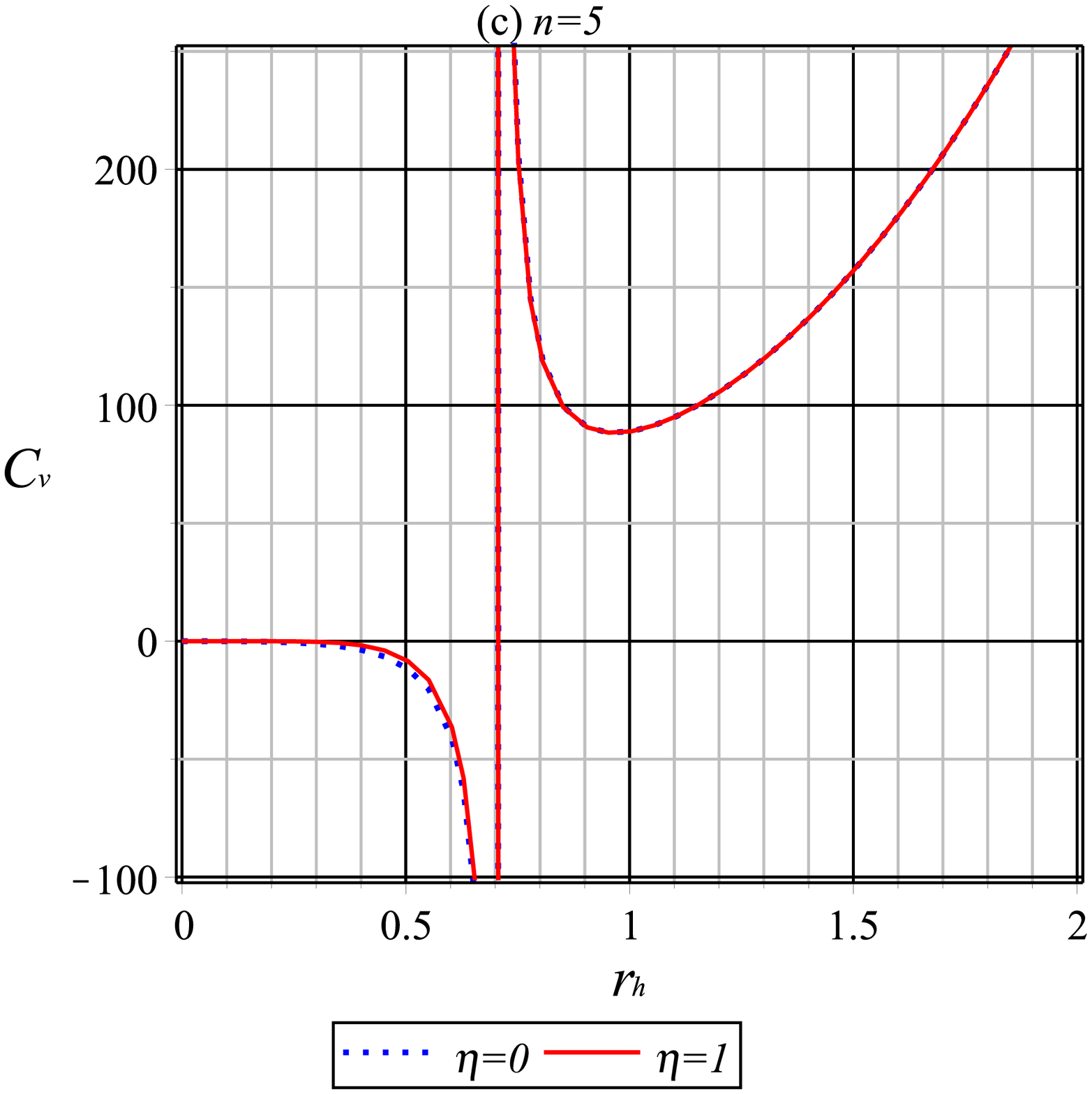}\includegraphics[width=50 mm]{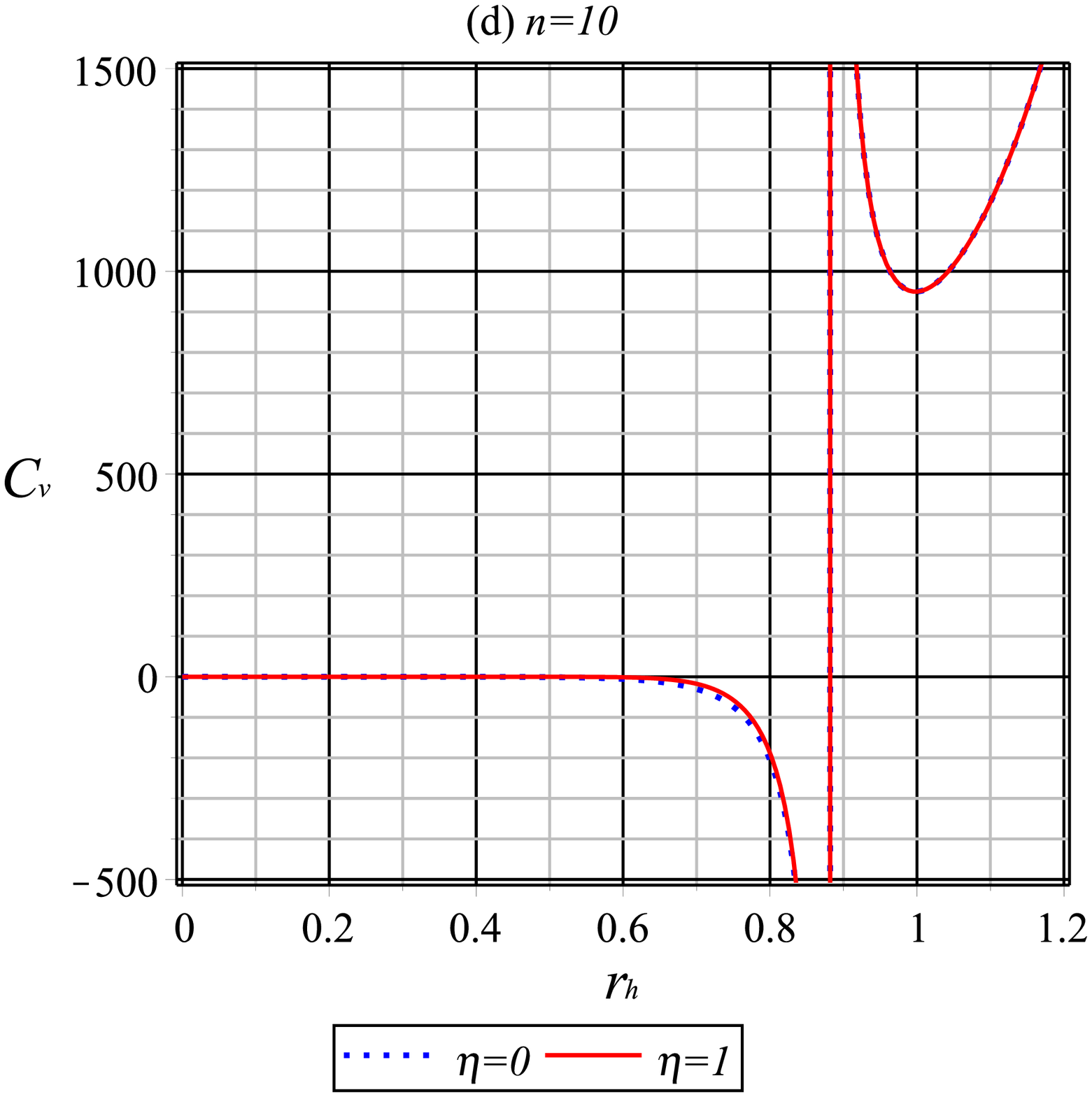}
\end{array}$
\end{center}
\caption{Specific heat of $n$-dimensional Schwarzschild-Tangherlini AdS black hole for $l=1$.}
\label{figCv}
\end{figure}

The first thing that we will analyze is the effect of these corrections to the stability of this system. This can be done by  studying   the effect of these non-pertubative quantum gravitational corrections on the specific heat of this system.
Thus, we will  use   the corrected entropy  $S$ given by Eq. (\ref{mod-ent}), to write the corrected  specific heat as
\begin{equation}\label{Cv}
C_{v}=T\left(\frac{\partial S}{\partial T}\right)_{V}
\end{equation}
This quantum gravitationally corrected specific heat of an AdS black hole can be explicitly written as
\begin{equation}
C_{v}=\left(\frac{\left(n-3\right)}{4\pi r_{+}}+\frac{\left(n-1\right)}{4\pi l^{2}}r_{+}\right)\left(\frac{\frac{\omega\left(n-2\right)}{2}r_{+}^{n-3}}{-\frac{\left(n-3\right)}{4\pi r_{+}^{2}}+\frac{\left(n-1\right)}{4\pi l^{2}}}\right)\left(1-\eta e^{-S_{0}}\right)
\end{equation}
In Fig. \ref{figCv},  we observe the behavior of specific heat in terms of  radius of the horizon,  and  we find that there occurs a phase transition for a  black hole with $n\geq4$, and such a black hole will  become  unstable at a sufficient small scales.  In Fig. \ref{figCv} (a),  we observe  that the  specific heat reduces  due to quantum corrections. However, this three-dimensional Ads black hole is stable  for all values of $r_h$. Thus, the stability of a AdS black hole depends on $n$ for such a black hole.

It is known that  for an  AdS black hole, the cosmological constant can be treated as the thermodynamic pressure, and a volume conjugate to it can be used to construct in an extended phase space  \cite{pv1, pv2, pv4, pv5}. So,  we can investigate  the effect of the non-perturbative quantum corrections \cite{3, 3aa, 3bb, ds12,  ds14} on the thermodynamics of an AdS black hole in this extended phase space.
The pressure and volume of this AdS black hole are given by  \cite{5, 51, 52, 54},
\begin{eqnarray}
P=\frac{(n-1)(n-2)}{16\pi l^{2}},
& \text{and} &
V=\frac{\omega}{n-1}r_{h}^{n-1}.
\end{eqnarray}
We can write the equation of state for this AdS black hole as  \cite{12a}
\begin{eqnarray}\label{EoS}
\frac{Pv}{T}=1-\frac{n-3}{(n-2)\pi}\frac{1}{Tv}, &\text{with}& v=\frac{4}{n-2}\left(\frac{(n-1)V}{
\omega_{n-2}}\right)^{\frac{1}{n-1}}
\end{eqnarray}
Comparing Eq. (\ref{EoS}) with the virial expansion, we observe
\begin{equation}\label{virial}
\frac{Pv}{T}=1+\frac{B(T)}{v}+\frac{C(T)}{v^{2}}+\frac{D(T)}{v^{3}}+\cdots,
\end{equation}
We can write  the non-vanishing virial coefficient as
\begin{equation}\label{B}
B(T)=-\frac{n-3}{(n-2)\pi T},
\end{equation}
here  $C(T)=D(T)=0$ (however, if we assume a four dimensional  charged black hole, then $D(T)\neq0$). In that case, Boyle temperature (the temperature where the virial coefficient $B(T)$ approaches zero) approaches infinity. It should be noted that this equation of state, given by Eq. (\ref{EoS}) is a special case of van der Waals equation of state
\begin{equation}\label{vdW}
Pv=\frac{T}{v-b}+\frac{a}{v^{2}},
\end{equation}
with $b=0$. For such an equation of state, we observe that  $a={n-3}/{(n-2)\pi}$.

Now we can analyze the behavior of  a very small black hole with a  quantum scale  radius. For such an AdS black hole, we can write
\begin{equation}
S\approx\frac{\omega}{2}(1-\eta)\left(\frac{n-2}{4}v\right)^{n-2}.
\end{equation}
The equation of state given by Eq. (\ref{EoS}), along with the condition for the  critical points, can be used to observe that
\begin{eqnarray}\label{cond}
\frac{\partial P}{\partial v}=0,\ &\text{and}&
\frac{\partial^{2} P}{\partial v^{2}}=0
\end{eqnarray}
This indicates that there is an absence of critical points, except at $v=0$.   Thus, we can calculate  the modification to stability of the system using the exponential corrections to the entropy from quantum gravitational effects.

\section{Corrected Thermodynamics}
 In this section, we will analyze the effects of these non-perturbative quantum gravitational corrections on other thermodynamic quantities of this system.
We can express the  quantum gravitationally corrected  internal energy of this AdS black hole as
\begin{eqnarray}
E&=&\int TdS \nonumber \\
 &=&\frac{\omega\left(n-2\right)}{8\pi}\left(r_{h}^{n-3}+\frac{r_{h}^{n-1}}{l^{2}}\right)+\eta\frac{\left(n-3\right)}{8\pi}2^{\frac{n-3}{n-2}}\omega^{\frac{1}{n-2}}\Gamma\left(\frac{n-3}{n-2},\frac{1}{2}r_{h}^{n-2}\omega\right)\nonumber\\
&+&\eta\frac{\left(n-1\right)}{8\pi l^{2}}2^{\frac{n-1}{n-2}}\omega^{-\frac{1}{n-2}}\Gamma\left(\frac{n-1}{n-2},\frac{1}{2}r_{h}^{n-2}\omega\right),
\end{eqnarray}
where $\Gamma (z,x)$ is the Gamma-function
\begin{equation}
\Gamma (z,x)=x^z e^{-x} \sum_{n=0}^{\infty} \frac{L_n^{(z)} (x)}{n+1}.
\end{equation}
Here $ L_n^{(z)} (x) $   the generalized Laguerre polynomial are defined as
\begin{eqnarray}
 L_0^{(z)} (x) &=& 0 \nonumber\\
 L_1^{(z)} (x) &=& 1+z-x \nonumber\\
 L_{k+1}^{(z)} (x) &=& \frac{(2k+1+z-x) L_k^{(z)} (x)-(k+z) L_{k-1}^{(z)} (x)}{k+1} \qquad  k \geq 1.
\end{eqnarray}

\begin{figure}
\begin{center}$
\begin{array}{cccc}
\includegraphics[width=50 mm]{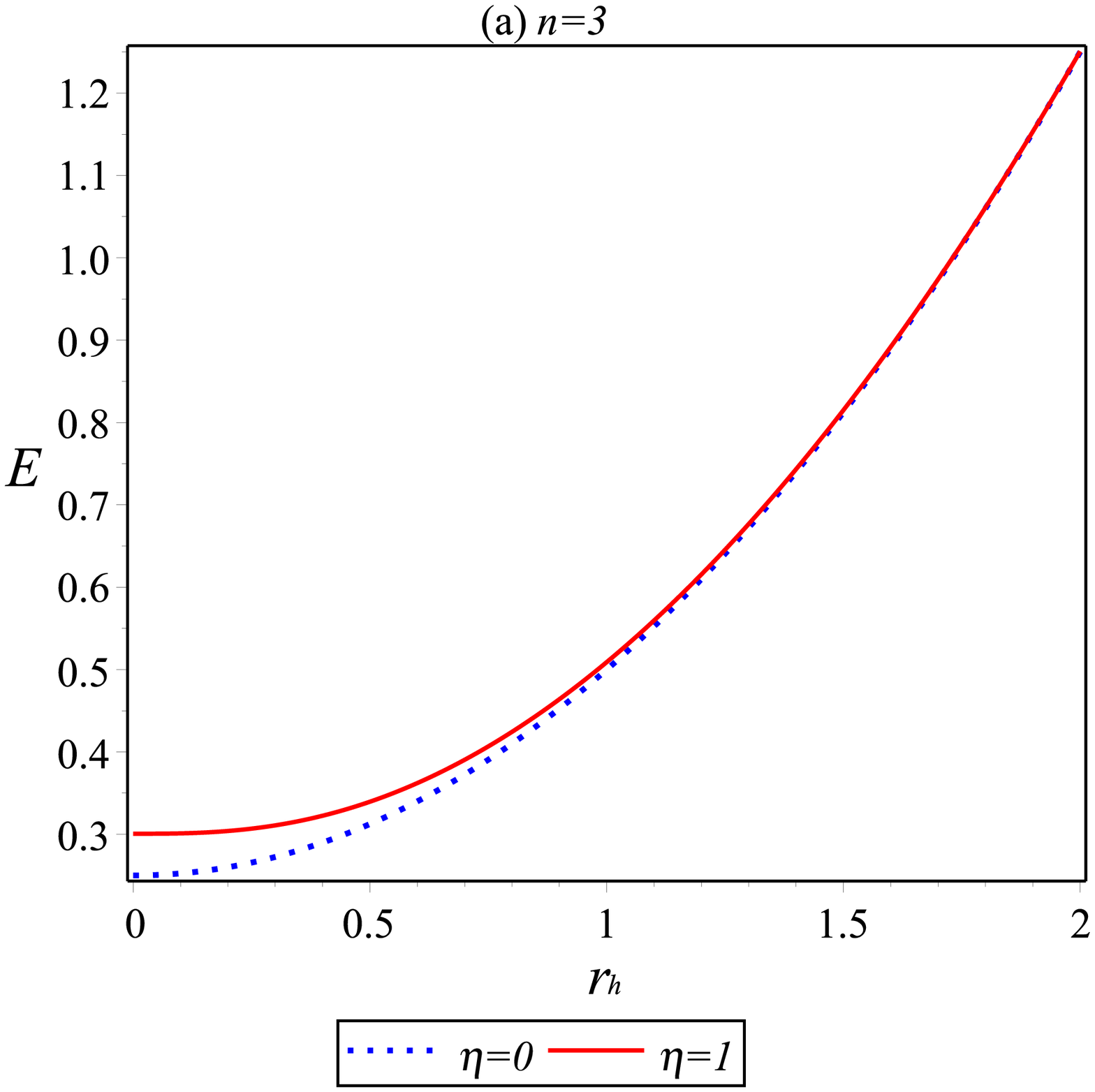}\includegraphics[width=50 mm]{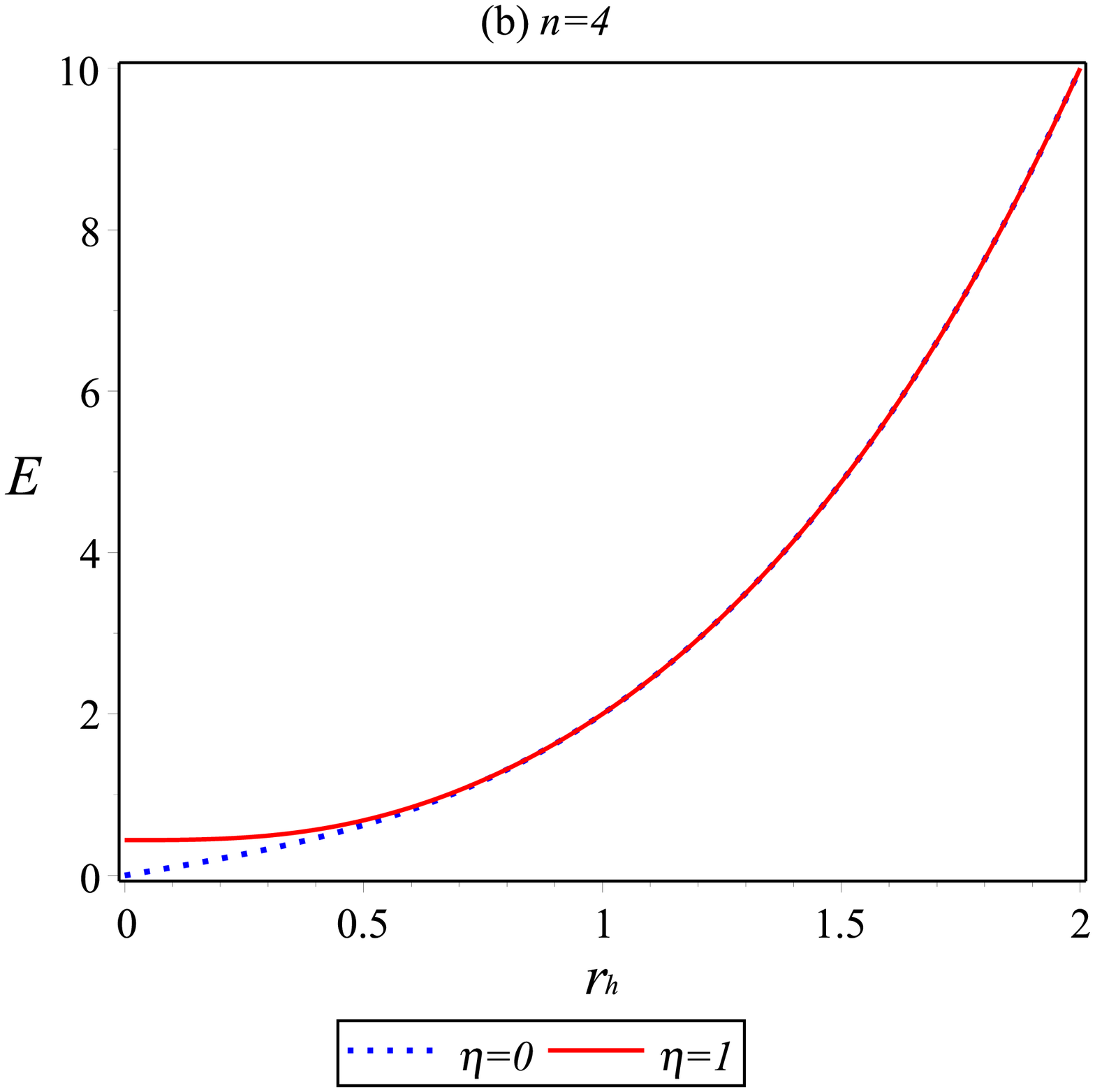}\\
\includegraphics[width=50 mm]{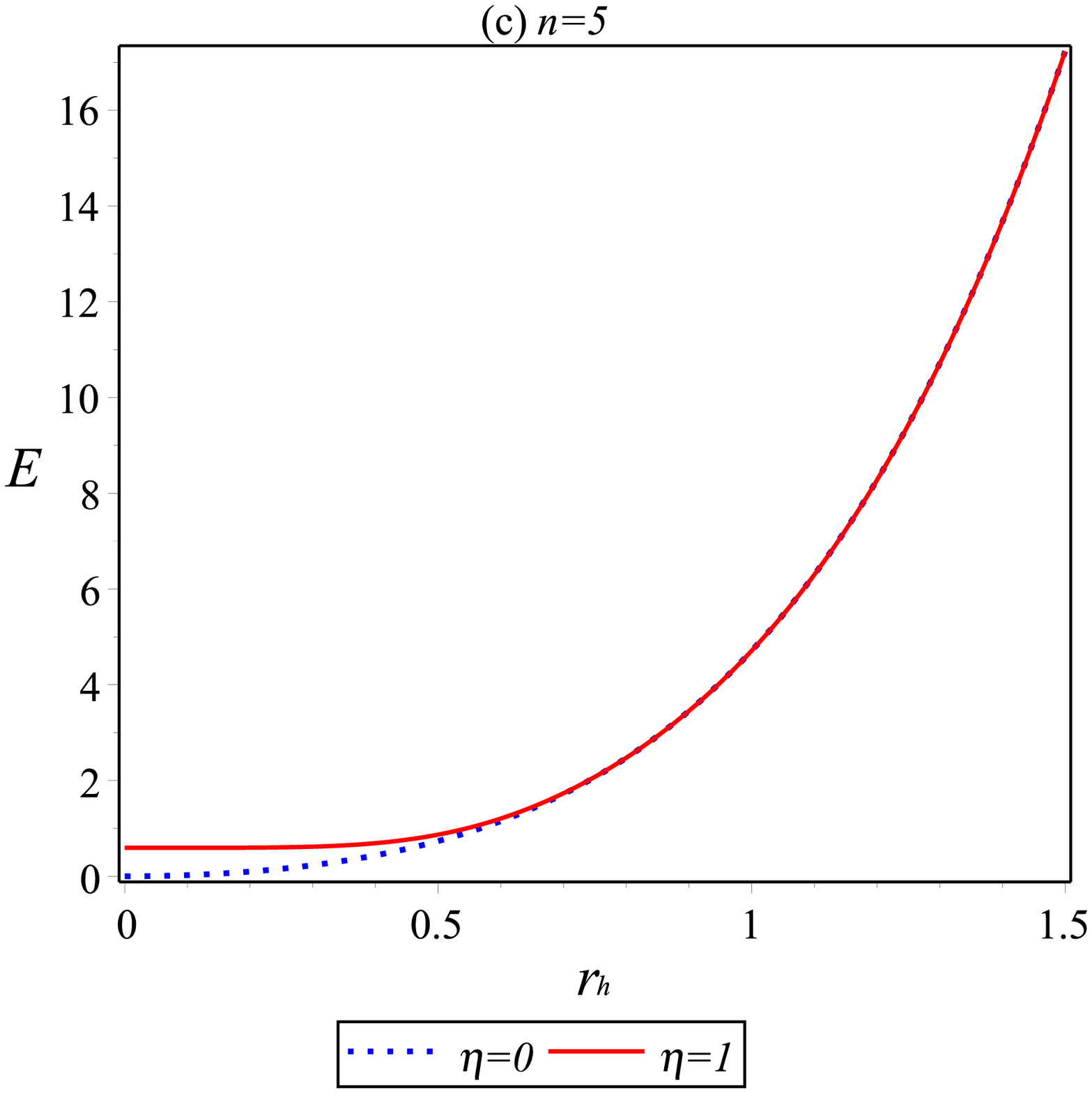}\includegraphics[width=50 mm]{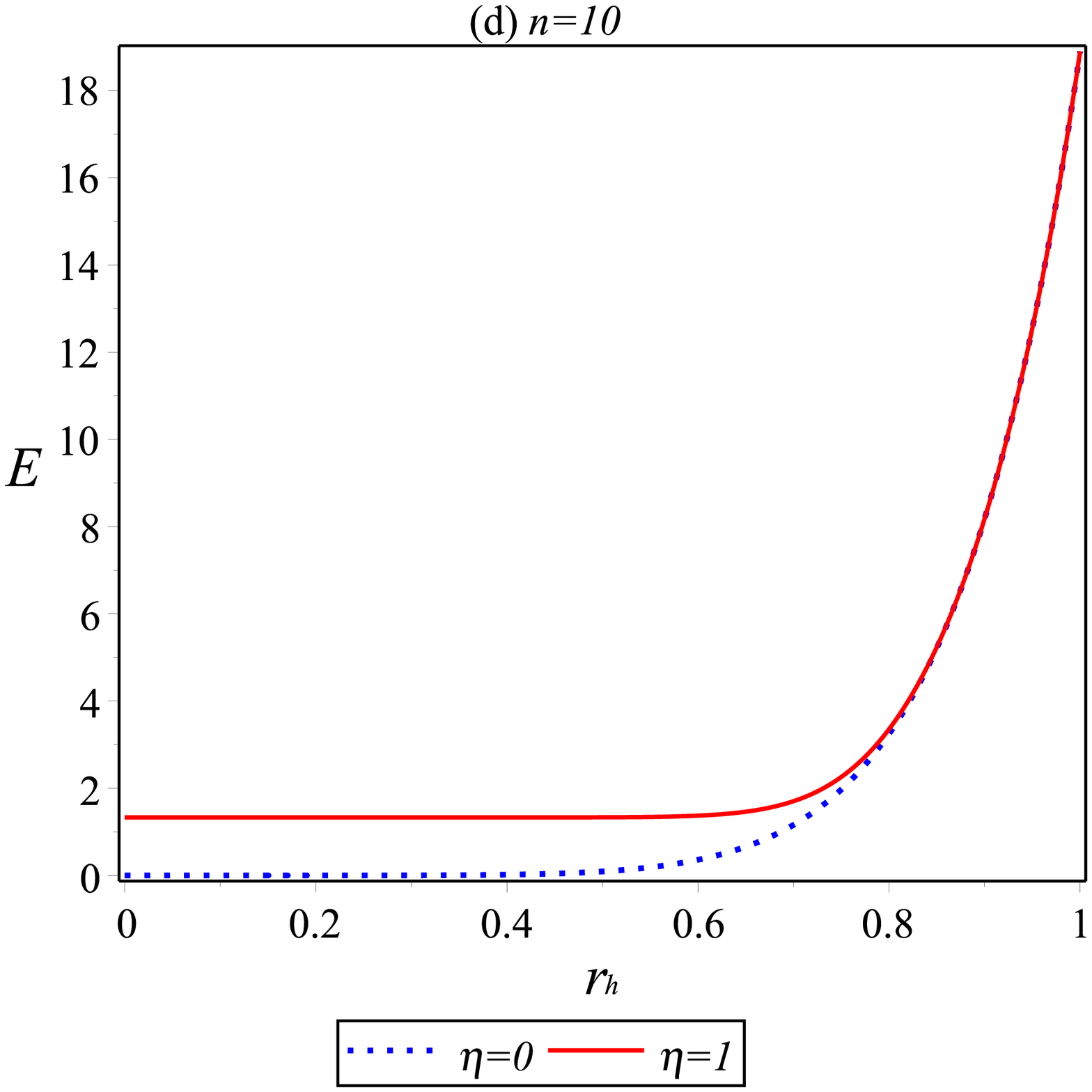}
\end{array}$
\end{center}
\caption{Internal energy of $n$-dimensional Schwarzschild-Tangherlini AdS black hole for $l=1$.}
\label{figE}
\end{figure}

In Fig.\ref{figE},  we plot the internal energy and observe that it increases with the increase in the radius of the  horizon. We also observe that the quantum gravitational corrections can be neglected for large black holes. However, at small quantum scales, these corrections change the internal energy of the AdS black hole.
We can obtain the  free energy for this AdS black hole using the quantum  corrected  internal energy, and entropy
$ F=E-TS $. Thus, for this AdS black hole the quantum corrected free energy can be written as
\begin{eqnarray}\label{F}
F&=&\frac{\omega\left((n-2)r_{h}(l^{2}r_{h}^{n-3}+r_{h}^{n-1})-r_{h}^{n-2}(l^{2}(n-3)+(n-1)r_{h}^{2})\right)}{8\pi r_{h}l^{2}}\nonumber\\
&+&\frac{\eta}{8\pi l^{2}}\left(2^{\frac{n-1}{n-2}}\omega^{-\frac{1}{n-2}}(n-1)\Gamma\left(\frac{n-1}{n-2},\frac{1}{2}r_{h}^{n-2}\omega\right)
+l^{2}2^{\frac{n-3}{n-2}}\omega^{\frac{1}{n-2}}(n-3)\Gamma\left(\frac{n-3}{n-2},\frac{1}{2}r_{h}^{n-2}\omega\right)\right)\nonumber\\
&-&\frac{\eta}{4\pi r_{h}l^{2}}(l^{2}(n-3)+(n-1)r_{h}^{2})\exp\left({-\frac{\omega r_{h}^{n-2}}{2}}\right).
\end{eqnarray}
We can now explicitly use quantum corrected free energy to calculate the quantum gravitationally corrections to Gibbs free energy. Thus, we
can  write  the quantum gravitationally corrected Gibbs free energy $G=F+PV$ as
\begin{eqnarray}
G&=&\frac{\omega\left(n-2\right)}{8\pi}\left(r_{h}^{n-3}+\frac{3r_{h}^{n-1}}{l^{2}}\right)
-\frac{\omega}{8\pi l^{2}}\left((n-1)r_{h}^{2}+(n-3)l^{2}\right)r_{h}^{n-3}\nonumber\\
&+&\frac{\eta}{8\pi l^{2}}\left(2^{\frac{n-1}{n-2}}\omega^{-\frac{1}{n-2}}(n-1)\Gamma\left(\frac{n-1}{n-2},\frac{1}{2}r_{h}^{n-2}\omega\right)
 +l^{2}2^{\frac{n-3}{n-2}}\omega^{\frac{1}{n-2}}(n-3)\Gamma\left(\frac{n-3}{n-2},\frac{1}{2}r_{h}^{n-2}\omega\right)\right)\nonumber\\
&-&\frac{\eta}{4\pi r_{h}l^{2}}(l^{2}(n-3)+(n-1)r_{h}^{2})\exp\left({-\frac{\omega r_{h}^{n-2}}{2}}\right).
\end{eqnarray}
\begin{figure}
\begin{center}$
\begin{array}{cccc}
\includegraphics[width=50 mm]{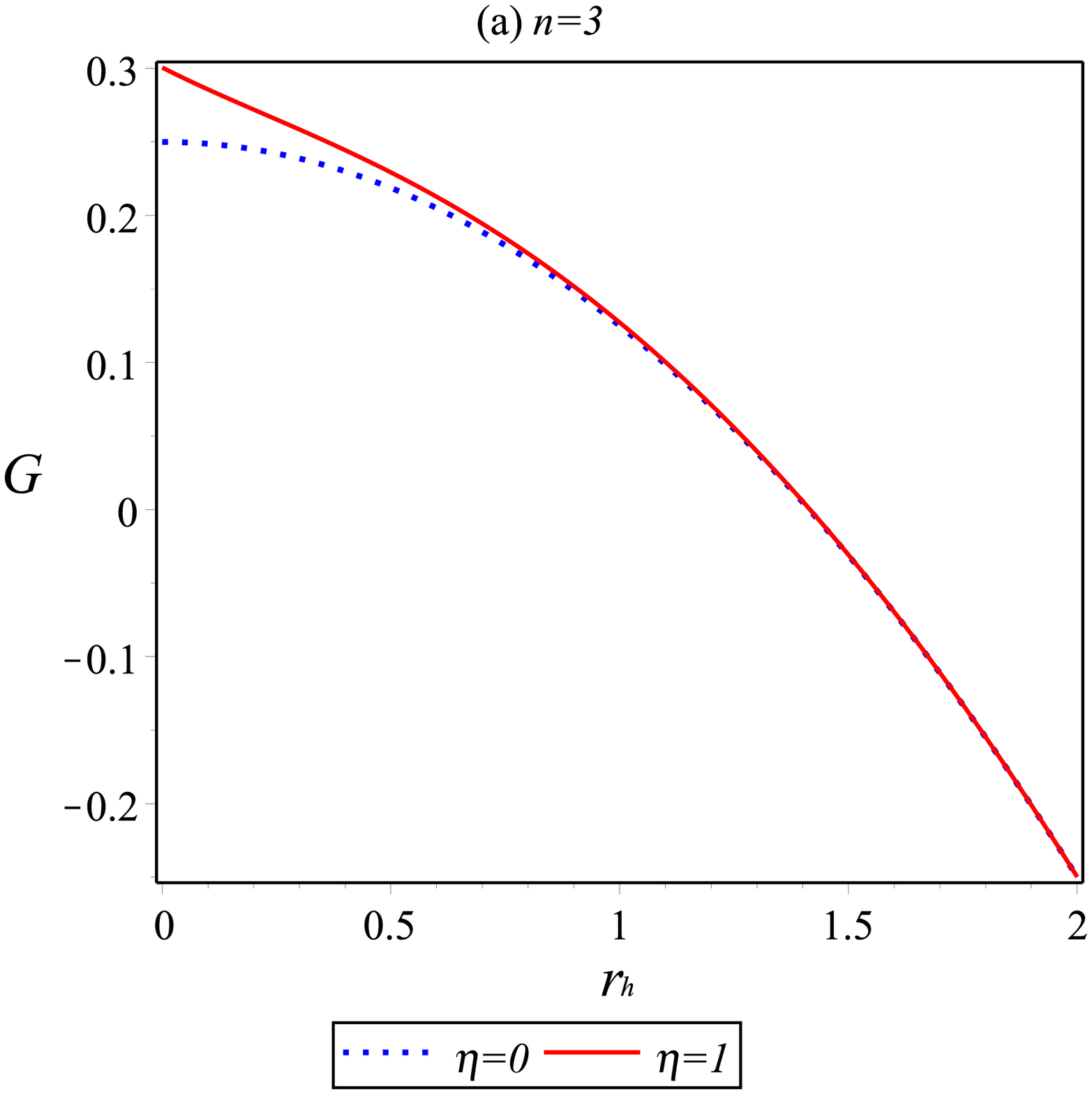}\includegraphics[width=50 mm]{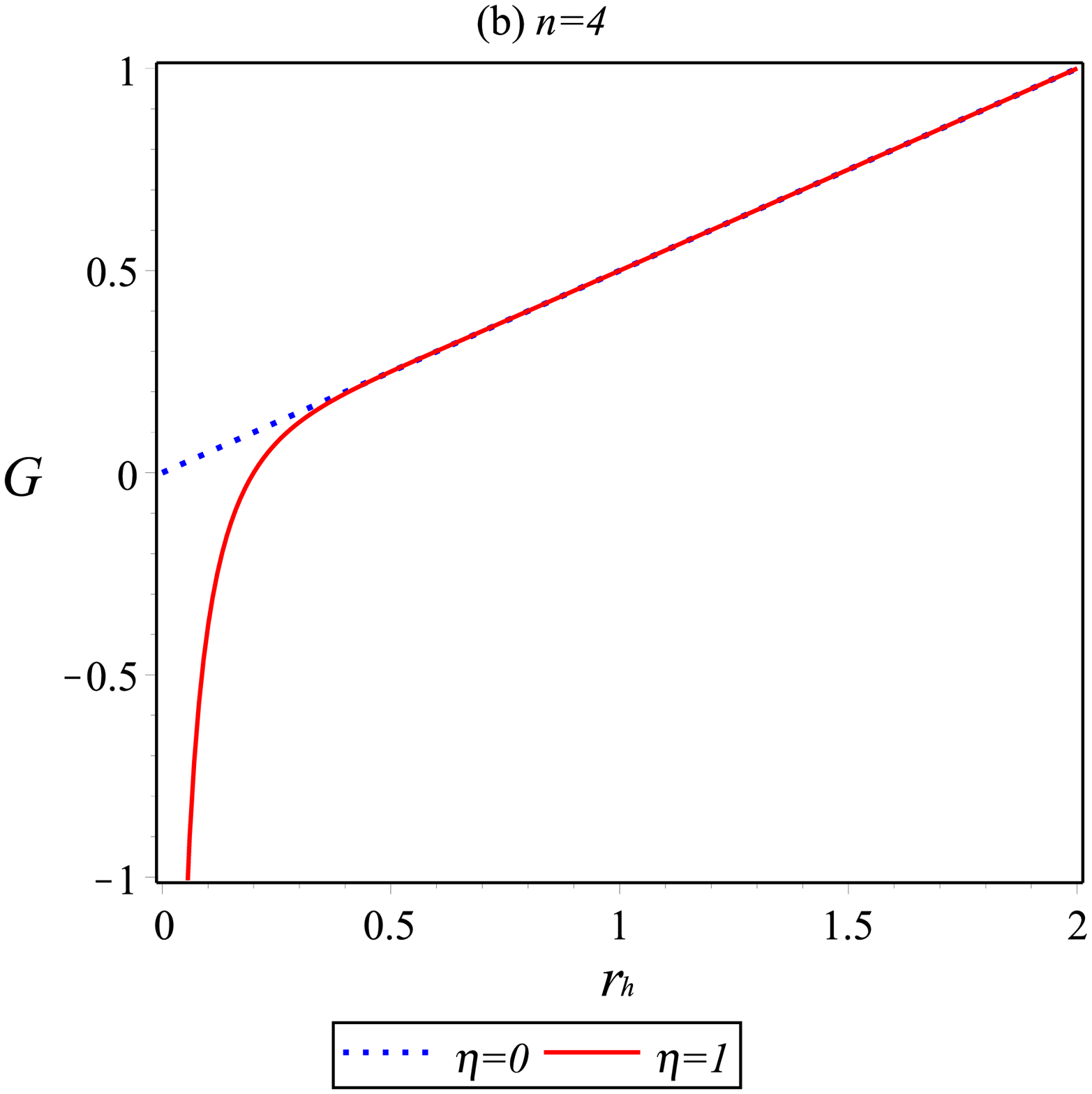}\\
\includegraphics[width=50 mm]{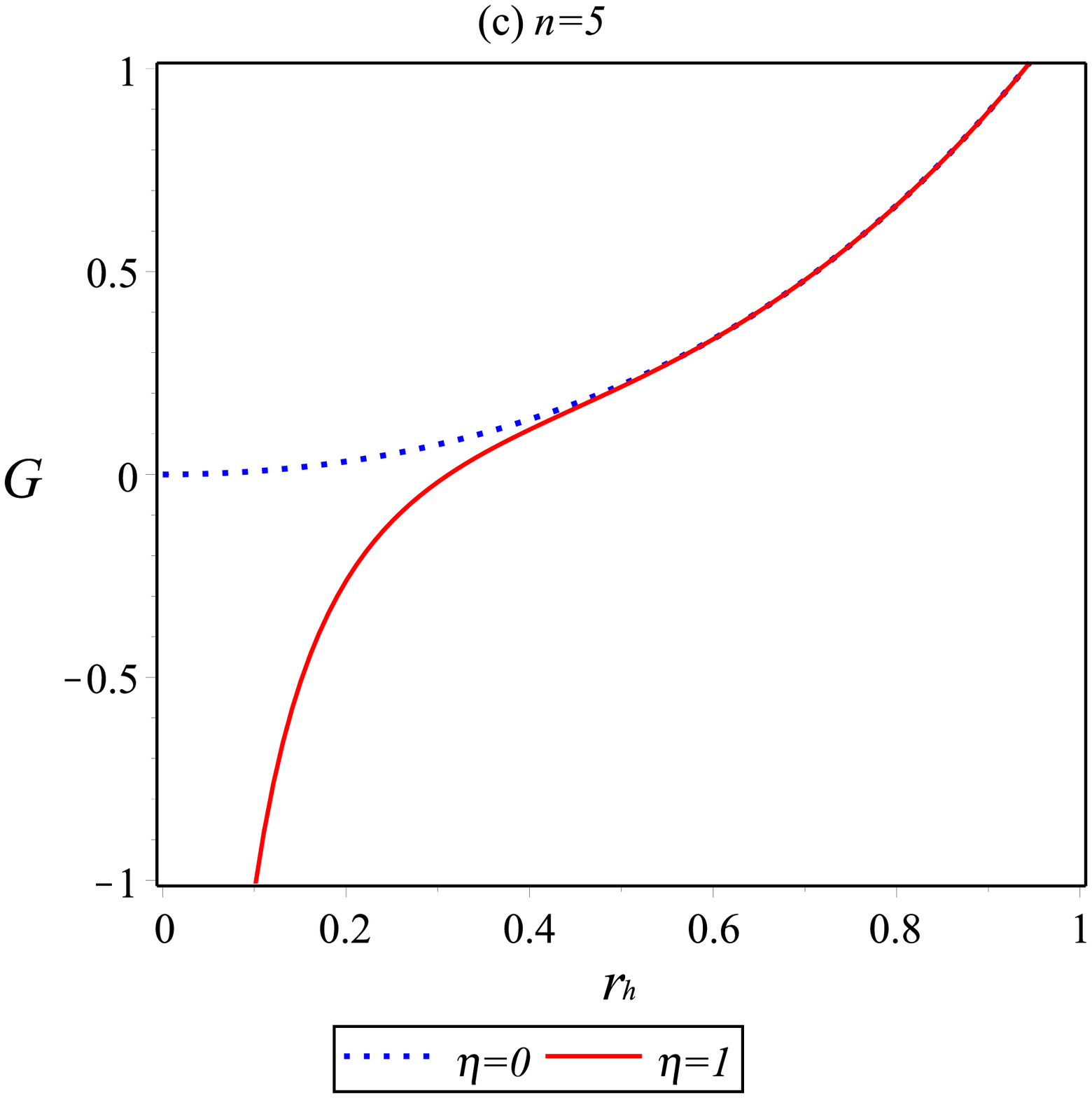}\includegraphics[width=50 mm]{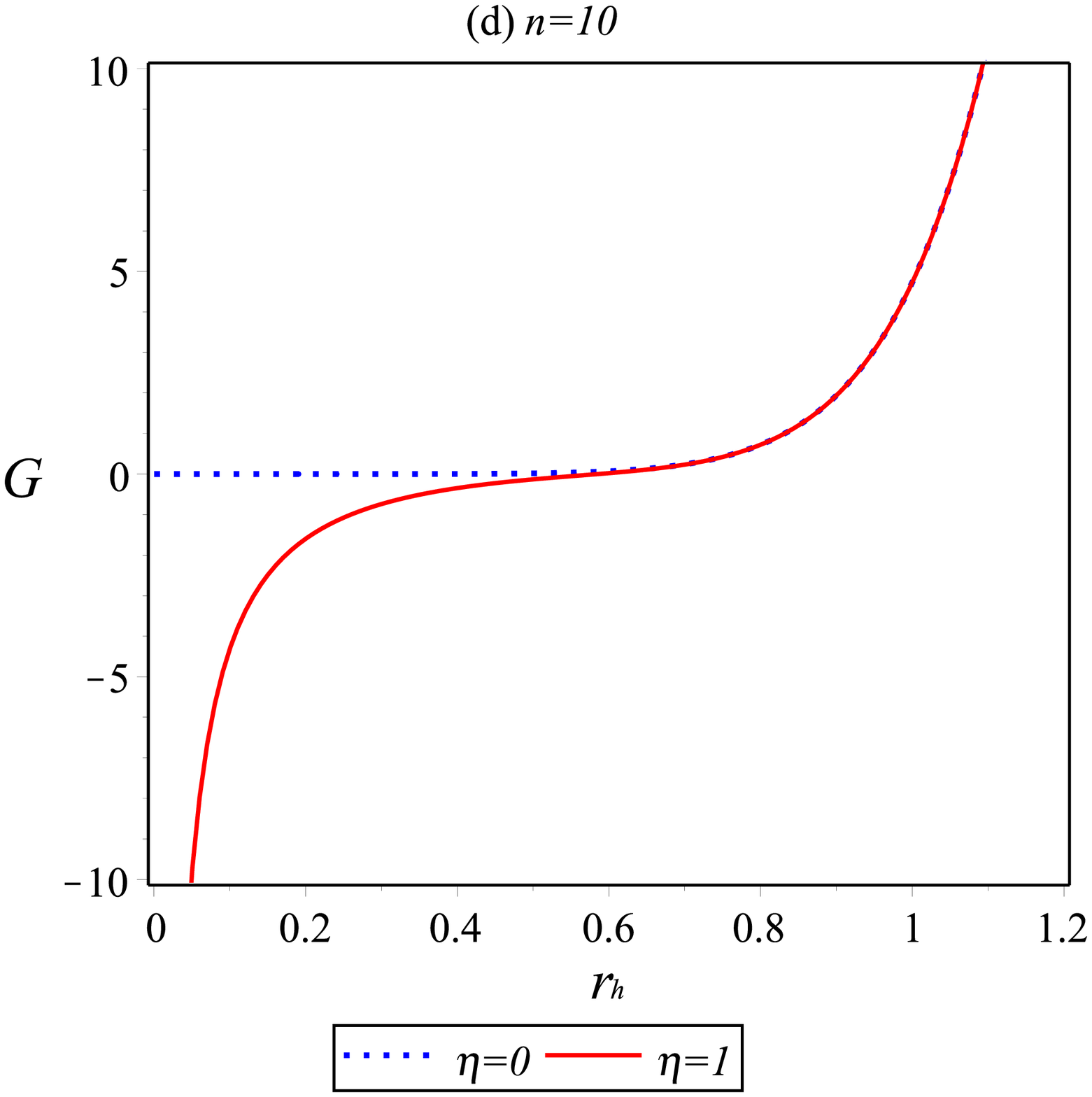}
\end{array}$
\end{center}
\caption{Gibbs free energy of $n$-dimensional Schwarzschild-Tangherlini AdS black hole for $l=1$.}
\label{figG}
\end{figure}

In Fig. \ref{figG}, we plot  the behavior of the quantum gravitationally corrected  Gibbs free energy.
It is again observed that the quantum gravitational corrections to the Gibbs free energy can be neglected at large scales. However, for quantum sized AdS black holes, the behavior of Gibbs free energy is significantly changed due to these quantum gravitational corrections. We also observe that the corrections to the Gibbs free energy depend on the dimensions of the AdS black hole.

\section{Quantum Work Distribution}
In this section, we analyze the quantum work distribution for an AdS black hole, as it evaporates between two states.
Thus, let  us assume that the microstates of the AdS black hole change from $\Omega_1$ to $\Omega_2$ during evaporation. Then the  partition function  of the black hole will  change  from  $Z_1 [\Omega_1]$ to $Z_2[\Omega_2]$. This will change different thermodynamic quantities between these two states.  As  the quantum gravitational corrections become important only at small  horizon radius,  we can express the difference in the corrected entropy of a quantum sized AdS black hole as
\begin{equation}
\Delta S=  \frac{\omega}{2}(1-\eta)(r_{h2}^{n-2}-r_{h1}^{n-2}).
\end{equation}
This difference between the  quantum gravitationally corrected entropy can be used to obtain  the difference between the internal energy of a black hole, as it evaporates between from  $\Omega_1$ to $\Omega_2$
\begin{eqnarray}
\Delta E&=&\frac{\omega\left(n-2\right)}{8\pi}\left(r_{h2}^{n-3}-r_{h1}^{n-3}+\frac{r_{h2}^{n-1}-r_{h1}^{n-1}}{l^{2}}\right)\nonumber\\
&+&\eta\frac{\left(n-3\right)}{8\pi}2^{\frac{n-3}{n-2}}\omega^{\frac{1}{n-2}}
\left[\Gamma\left(\frac{n-3}{n-2},\frac{1}{2}r_{h2}^{n-2}\omega\right)-\Gamma\left(\frac{n-3}{n-2},\frac{1}{2}r_{h1}^{n-2}\omega\right)\right]\nonumber\\
&+&\eta\frac{\left(n-1\right)}{8\pi l^{2}}2^{\frac{n-1}{n-2}}\omega^{-\frac{1}{n-2}}
\left[\Gamma\left(\frac{n-1}{n-2},\frac{1}{2}r_{h2}^{n-2}\omega\right)-\Gamma\left(\frac{n-1}{n-2},\frac{1}{2}r_{h1}^{n-2}\omega\right)\right].
\end{eqnarray}
These corrections  to the internal energy will produce     quantum gravitational corrections to the Hawking radiation, for quantum sized AdS black holes. We can denote  this  total  heat obtained   from quantum gravitationally  corrected  Hawking radiation by  $Q$.
Now at such scales, we cannot neglect the effects of quantum work, as the size of the black hole reduces. So, we obtain finite quantum work  as the black hole evaporates from $\Omega_1$ to $\Omega_2$. If we denote the average  quantum work by $\langle W \rangle$, we can write
\begin{equation}
\Delta E =  Q - \langle W \rangle
\end{equation}
Now let us assume that an AdS black hole with a partition function  $Z_1[\Omega_1]$   evaporates to an AdS black hole with a partition function $Z_2[\Omega_2]$.
The term  $Z_2/Z_1$ can be related to the quantum work distribution, using  the Jarzynski equality    \cite{eq12, eq14}
\begin{equation}
\langle \exp {-\beta W} \rangle \nonumber=\frac{Z_2}{Z_1}
\end{equation}
We can also express  the relative weights of the partition function to the difference between the  equilibrium  free energies as
$
\exp {\beta \Delta F} = {Z_2}/{Z_1}
$.
Thus, we can use the   Jarzynski equality   \cite{eq12, eq14} to express the    quantum work in terms of  difference of the equilibrium  free energies of these two states  for an AdS black hole
\begin{equation}
\langle \exp {-\beta W} \rangle = \exp {\beta \Delta F}
\end{equation}

We can use this quantum corrected free energy to calculate the average quantum work between these two black hole states.
It is possible to use the Jensen inequality to relate  the average of the exponential of quantum work to the exponential of the average of quantum work as as
$ \exp {\langle -\beta {W} \rangle } \leq \langle \exp {-\beta W} \rangle
$.  Using this inequality,  we can express the  quantum work distribution during the evaporation of an AdS black hole as
\begin{eqnarray}
 \langle  W \rangle &\approx&-\frac{\omega\left((n-2)r_{h2}(l^{2}r_{h2}^{n-3}+r_{h2}^{n-1})-r_{h2}^{n-2}(l^{2}(n-3)+(n-1)r_{h2}^{2})\right)}{8\pi r_{h2}l^{2}}\nonumber\\
&+&\frac{\omega\left((n-2)r_{h1}(l^{2}r_{h1}^{n-3}+r_{h1}^{n-1})-r_{h}1^{n-2}(l^{2}(n-3)+(n-1)r_{h1}^{2})\right)}{8\pi r_{h1}l^{2}}\nonumber\\
&-&\frac{\eta}{8\pi l^{2}}\left(2^{\frac{n-1}{n-2}}\omega^{-\frac{1}{n-2}}(n-1)\Gamma\left(\frac{n-1}{n-2},\frac{1}{2}r_{h2}^{n-2}\omega\right)
+l^{2}2^{\frac{n-3}{n-2}}\omega^{\frac{1}{n-2}}(n-3)\Gamma\left(\frac{n-3}{n-2},\frac{1}{2}r_{h2}^{n-2}\omega\right)\right)\nonumber\\
&+&\frac{\eta}{8\pi l^{2}}\left(2^{\frac{n-1}{n-2}}\omega^{-\frac{1}{n-2}}(n-1)\Gamma\left(\frac{n-1}{n-2},\frac{1}{2}r_{h1}^{n-2}\omega\right)
+l^{2}2^{\frac{n-3}{n-2}}\omega^{\frac{1}{n-2}}(n-3)\Gamma\left(\frac{n-3}{n-2},\frac{1}{2}r_{h1}^{n-2}\omega\right)\right)\nonumber\\
&-&\frac{2\eta}{8\pi r_{h2}l^{2}}(l^{2}(n-3)+(n-1)r_{h2}^{2})\left(1-\frac{\omega r_{h2}^{n-2}}{2}\right)\nonumber\\
&+&\frac{2\eta}{8\pi r_{h1}l^{2}}(l^{2}(n-3)+(n-1)r_{h1}^{2})\left(1-\frac{\omega r_{h1}^{n-2}}{2}\right)
\end{eqnarray}
\begin{figure}
\begin{center}$
\begin{array}{cccc}
\includegraphics[width=70 mm]{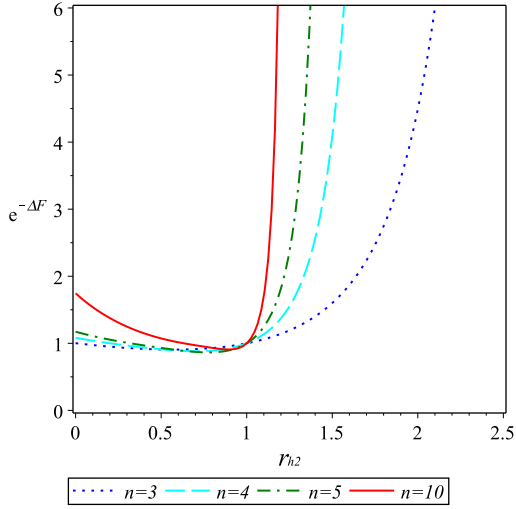}
\end{array}$
\end{center}
\caption{$e^{-\Delta F}$ for $n$-dimensional Schwarzschild-Tangherlini AdS black hole, with $r_{h1}=1$, $\eta=1$ and $l=1$.}
\label{fig-exp-delta}
\end{figure}
This is plotted  in Fig. \ref{fig-exp-delta}. We can use this expression to express the quantum work in terms of the partition functions $Z_1$ and $Z_2$ as $\langle e^{-\beta W}\rangle = Z_2/Z_1$ \cite{6b}. So, the relative weights of the partition functions of a black hole depends on the quantum work done between those two states. This is expected as the quantum work is dependent on the difference between equilibrium free energies between these two states, which also depends on the microstates of the black hole. Now,  as the black hole evaporates, it emits Hawking radiation. However, we have to also consider average  quantum work between these two states. The terms related to quantum work becomes important only at small scales, and so we had to  consider the effects of  quantum gravitational corrections on it. These corrections modified the expression for equilibrium free energies, and this in turn modified the  expression for quantum work. We had to use this modified expression for free energies to analyze the effect of quantum gravitational corrections on quantum work distribution.

\section{Conclusion }

In this paper, we have studied the thermodynamics of a quantum sized Schwarzschild-Tangherlini AdS black hole in higher dimension. We   used  quantum gravitationally  corrected  entropy to analyze the corrections to the  thermodynamic behavior of such a black hole. We considered the  effect of such  non-perturbative quantum gravitational  corrections on the stability of such a black hole. This was done by analyzing  quantum  corrections to the  specific heat  for this AdS black hole. The effect of such corrections on the Gibbs free energy was also investigated.  Finally, we observed that at such a small scale, we cannot neglect the average quantum work  between two black hole states.
The difference in free energies between two states was used to obtain    quantum work distribution. This was done using  the Jarzynski equality.  We also discussed the  relation between   quantum work distribution and relative weights of  the partition for an evaporating AdS black hole.

It seems important that the effect of such  non-perturbative quantum gravitational correction  should be investigated for other quantum sized black holes.  It would be interesting to analyze such corrections to extra-dimensional black objects, such as black  strings.  This formalism can be used to analyze the corrections to the    thermodynamics of black strings. We can also calculate the average  quantum work  for the black strings using the difference between free energies.  In the context of singularity theorems, we can use such quantum gravitational corrections to obtain   modified quantum Raychaudhuri equation for those effective geometries. It is expected that the geometrical flow could be effected by such  non-perturbative quantum gravitational  corrections.  It would be interesting to investigate such effects.
%%%%%%%%%%%%%%%%%%%%%%%%%%%%%%%%%%%%%%%%%%%%%%%%%%%%%%%%%%%%%%%%%%%%%
%%%%%%%%%%%%%%%%%%%%%%%%%%%%%%%%%%%%%%%%%%%%%%%%%%%%%%%%%%%%%%%%%5

\end{document}